\documentstyle[art12]{article}
\catcode`\@=11
\catcode`\@12
\begin{document}
\pagestyle{myheadings}
\typeout{}
\newcommand{\eqn}[1]{(\ref{eq:#1})}
\newcommand{\ben}{\begin{equation}}
\newcommand{\een}{\end{equation}}
\newcommand{\bea}{\begin{eqnarray}}
\newcommand{\eea}{\end{eqnarray}}
\newcommand{\nn}{\nonumber \\ }
\newcommand{\bdm}{\begin{displaymath}}
\newcommand{\edm}{\end{displaymath}}
\newcommand{\pa}{\partial }
\newcommand{\hf}{\frac{1}{2}}
\newcommand{\slr}{sl(2/1;{\bf R})}
\newcommand{\slc}{sl(2/1;{\bf C})}
\newcommand{\hslr}{\hat{sl}(2/1;{\bf R})}
\newcommand{\hslc}{\hat{sl}(2/1;{\bf C})}
\newcommand{\Slr}{SL(2/1;{\bf R})}
\newcommand{\rt}{\sqrt{2k}}
\newcommand{\irt}{\frac{1}{\sqrt{2k}}}
\newcommand{\frt}{\sqrt{\frac{k}{2}}}
\newcommand{\ifrt}{\sqrt{\frac{2}{k}}}
\newcommand{\iap}{i\alpha_+}
\newcommand{\pap}{\psi'_1}
\newcommand{\pbp}{\psi'_2}
\newcommand{\psa}{\psi_1}
\newcommand{\psb}{\psi_2}
\newcommand{\papd}{\psi^{'\dagger}_1}
\newcommand{\pbpd}{\psi^{'\dagger}_2}
\newcommand{\fm}{\phi_-}
\newcommand{\fp}{\phi_+}
\newcommand{\tpap}{\tilde{\psi}'_1}
\newcommand{\tpbp}{\tilde{\psi}'_2}
\newcommand{\tpsa}{\tilde{\psi}_1}
\newcommand{\tpsb}{\tilde{\psi}_2}
\newcommand{\tpapd}{\tilde{\psi}^{'\dagger}_1}
\newcommand{\tpbpd}{\tilde{\psi}^{'\dagger}_2}
\newcommand{\tfm}{\tilde{\phi}_-}
\newcommand{\tfp}{\tilde{\phi}_+}
\newcommand{\Bp}{\exp[\tilde{\phi}\cdot{\bf h}]}
\newcommand{\Bpi}{\exp[-\tilde{\phi}\cdot{\bf h}]}
\parskip 2ex

\begin{titlepage}
\begin{flushright}
DTP/96/11\\
May 1996\\
\end{flushright}
\vspace{1cm}
\begin{center}
{\Large\bf Free field representations for the affine 
superalgebra $\widehat{sl(2/1)}$ and noncritical $N=2$ strings.}\\
\vspace{1cm}
{\large
P. Bowcock, R-L. K. Koktava and A. Taormina}\\
\vspace{0.5cm}
{\it Department of Mathematical Sciences, University of Durham,\\
Durham DH1~3LE, England} \\
\vspace{0.5cm}
\end{center}
\begin{abstract}
{}Free field representations of the affine superalgebra $A(1,0)^{(1)}$ 
at level $k$ are needed in the description of the noncritical $N=2$ string.
The superalgebra admits two inequivalent choices of simple roots. We give
the Wakimoto representations corresponding to each of these and derive
the relation between the two at the quantum level.
\end{abstract}
\vskip 5cm
e-mail addresses: Peter.Bowcock@durham.ac.uk, Anne.Taormina@durham.ac.uk
\end{titlepage}

In recent years, noncritical strings have been the focus of intense activity.
Most encouragement came from the nonperturbative definition of string theory
in space-time dimension $d < 1$ in the context of matrix models.
Although less powerful, the continuum approach, which involves the quantisation 
of the Liouville theory, gives results which are in agreement with those
obtained in matrix models, on the scaling behaviour of correlators for instance
\cite{goulian}. A generalisation of these ideas
roots are shown to be related by nonlinear canonical field 
transformations, both at the classical and at the quantum level. The
to supersymmetric strings however is easier
in the continuum. Some effort has been put in the study of $N=1$ and $N=2$
noncritical superstrings, but no clear picture has emerged so far as how useful
they might be, in particular in extracting nonperturbative information
\cite{distler,fy2,abdalla1,anton,abdalla2}. However, the $N=2$ 
noncritical string possesses interesting features and technical 
challenges. In particular, as emphasized in \cite{distler},
the $N=2$ noncritical string is not confined to the regime of weak 
gravity, i.e. the phase transition point between weak and strong gravity 
regimes is not of the same nature as in the $N=0,1$ cases. This absence of 
barrier in the central charge is source of complications, but also the hope 
of some new physics.

In this letter, we provide some of the algebraic background required 
to describe the space of physical states of the noncritical $N=2$ string, 
from the point of view developed
in \cite{yankiel,huyu,fy1,fy2} for the $N=0$ and $N=1$ cases. It is argued 
there
that gauged
$G/G$ Wess-Zumino-Novikov-Witten (WZNW) models, with $G$ a Lie (super)group, 
are promising tools for the study of noncritical (super)strings. 
In particular,
the $\Slr /\Slr $ topological quantum field theory obtained by gauging 
the
anomaly free diagonal subgroup $\Slr$ of the global $\Slr _L \times \Slr _R$
symmetry of the WZNW model appears to be 
intimately related to the noncritical charged fermionic string, which 
is the prototype of $N=2$ supergravity in two dimensions.  
A comparison of the ghost content of the two theories strongly 
suggests that the $N=2$ noncritical string is equivalent to the 
tensor product of a $\em twisted$ $\Slr /\Slr $ WZNW model with 
the topological theory of a spin $1/2$ system \cite{us}. 

{}For the bosonic and fermionic noncritical 
strings, the gauged WZNW action is based on the Lie group 
$SL(2;{\bf R})$ and the
Lie supergroup $Osp(1/2;{\bf R})$ respectively. 
It is also plausible that $W_N$ strings are related to the 
$SL(N;{\bf R})/SL(N,{\bf R})$ WZNW model
\cite{yankiel2,sadov}. It is however only when a one-to-one correspondence 
between the physical states and equivalence of the correlation functions 
of the two theories are established that one can view the twisted $G/G$ 
model as the topological version of the corresponding noncritical string 
theory. For the bosonic string, 
the recent derivation of conformal blocks for admissible representations
of $\widehat{sl(2;{\bf R})}$ is a major step in this direction \cite{jens}.

The physical states of the $\Slr /\Slr $ theory will be obtained in a 
forthcoming publication as elements of the cohomology of the BRST 
charge \cite{us}. The procedure we follow is by now quite standard 
\cite{bouw,yankiel,fy1,fy2}. The partition function of the 
$\Slr /\Slr $ theory splits in three sectors : a level $k$  
and a level $-(k+2)$ WZNW models based on $\Slr$
as well as  a system of four fermionic ghosts 
$(b_a,c^a), a=\pm , 3,4$ and four bosonic ghosts 
$(\beta_{\alpha},\gamma^{\alpha}),
(\beta'_{\alpha}, \gamma'_{\alpha}),\alpha =\pm \hf$
corresponding to the four even (resp. odd) generators of $\Slr$ 
\cite{PW,ks,gk,yankiel}.

The cohomology is calculated on the space 
${\cal{F}} _k \otimes {\cal{F}}_{-(k+2)}
\otimes {\cal {F}}_2$ where ${\cal{F}}_k$ denotes the space of 
irreducible representations of $\hslr_k$, while 
${\cal{F}}_{-(k+2)}$ and ${\cal{F}}_2$ denote the Fock spaces of 
the level $-(k+2)$ and ghosts sectors respectively. As a first step, 
one calculates the cohomology on the whole Fock space, using a 
free field  representation of $\hslr$ and its dual. These are the 
Wakimoto modules presented below.  Because of the non-unique 
interpretation of the Dynkin diagram for the Lie superalgebra $A(1,0)$ (the
complexification $\slc$ of $\slr$ in Kac's notations \cite{kac77}), 
one can associate two different free field representations 
to the two Weyl inequivalent choices of simple roots. The highly non 
linear relations between the free fields of these two representations 
are worked out in detail,  enabling one to obtain the physical states
of the noncritical $N=2$ string unambiguously.

In a second step, one must pass from the cohomology on the Fock space 
to the irreducible representations of $A(1,0)^{(1)} \sim \hslc$ at fractional 
level $k$.
Within this class of representations, only those called ${\em admissible}$
\cite{kw} are of interest for the problem at end. They are irreducible,
usually nonintegrable, and their characters transform as finite representations of the modular group. A detailed analysis of the representation theory of
$\hslc$ is given elsewhere \cite{us2}.

The set $\cal{M}$ of $3 \times 3$ matrices with real entries $m_{ij}$
whose diagonal elements satisfy the super-tracelessness condition
\ben
m_{11}+m_{22}-m_{33}=0
\een
forms, with the standard laws of matrix addition and multiplication,
the real Lie superalgebra $\slr$.
Any matrix ${\bf m} \in \cal{M}$ can
be expressed as a real linear combination of eight basis matrices

\bea
{\bf m}&=&m_{11}{\bf h_1}+m_{22}{\bf h_2}+m_{12}{\bf e_{\alpha_1 +\alpha_2}} 
       +m_{21}{\bf e_{-(\alpha_1 +\alpha_2)}}\nn
       &+&m_{32}{\bf e_{\alpha_1}}+m_{23}{\bf e_{-\alpha_1}}
       + m_{13}{\bf e_{\alpha_2}}+m_{31}{\bf e_{-\alpha_2}}
\eea
with
\bea
{\bf h}_1 = \left(
\begin{array}{cccc}
1 & 0 & \vrule & 0 \\
0 & 0 & \vrule & 0 \\
\hline 
0 &  0 & \vrule & 1 \\
\end{array}
\right ),&&
{\bf h}_2 = \left(
\begin{array}{cccc}
0 & 0 & \vrule & 0 \\
0 & 1 & \vrule & 0 \\
\hline 
0 & 0 & \vrule & 1 \\
\end{array}
\right ),\nonumber \\
{\bf e}_{\alpha_1+\alpha_2} = \left(
\begin{array}{cccc}
0 & 1 & \vrule & 0 \\
0 & 0 & \vrule & 0 \\
\hline 
0 & 0 & \vrule & 0 \\
\end{array}
\right ),&&
{\bf e}_{-(\alpha_1+\alpha_2)}=\left(
\begin{array}{cccc}
0 & 0 & \vrule & 0 \\
1 & 0 & \vrule & 0 \\
\hline 
0 & 0 & \vrule & 0 \\
\end{array}
\right ),\nonumber \\
{\bf e}_{\alpha_1}=\left(
\begin{array}{cccc}
0 & 0 & \vrule & 0\\
0 & 0 & \vrule & 0 \\
\hline 
0 & 1 & \vrule & 0 \\
\end{array}
\right ),&&
{\bf e}_{-\alpha_1}=\left(
\begin{array}{cccc}
0 & 0 & \vrule & 0 \\
0 & 0 & \vrule & 1 \\
\hline 
0 & 0 & \vrule & 0 \\
\end{array}
\right ),\nonumber \\
{\bf e}_{\alpha_2}=\left(
\begin{array}{cccc}
0 & 0 & \vrule & 1 \\
0 & 0 & \vrule & 0 \\
\hline 
0 & 0 & \vrule & 0 \\
\end{array}
\right ),&&
{\bf e}_{-\alpha_2}=\left(
\begin{array}{cccc}
0 & 0 & \vrule & 0 \\
0 & 0 & \vrule & 0 \\
\hline 
1 & 0 & \vrule & 0 \\
\end{array}
\right ).
\label{eq:fr}\eea
{}From this
fundamental 3-dimensional representation of $\slr$, one can write down
the (anti)-commutation relations obeyed by its four bosonic generators
${H_{\pm},E_{\pm(\alpha_1+\alpha_2)}}$ (corresponding to the even
basis matrices ${\bf h_{\pm}}={\bf h_1} \pm {\bf h_2}, {\bf
e_{\pm(\alpha_1 +\alpha_2)}}$) and its four fermionic generators
$E_{\pm \alpha_1}, E_{\pm \alpha_2}$ (corresponding to the odd basis
matrices),
\bea
&&[E_{\alpha_1+\alpha_2},E_{-(\alpha_1+\alpha_2)}]=H_1-H_2~~~,~~~
[H_1-H_2,E_{\pm(\alpha_1+\alpha_2)}]=\pm 2E_{\pm(\alpha_1+\alpha_2)},\nn
&&[E_{\pm(\alpha_1+\alpha_2)},E_{\mp \alpha_1}]=\pm E_{\pm \alpha_2}~~~,~~~
[E_{\pm(\alpha_1+\alpha_2)},E_{\mp \alpha_2}]=\mp E_{\pm \alpha_1},\nn
&&[H_1-H_2,E_{\pm \alpha_1}]=\pm  E_{\pm \alpha_1}~~~,~~~
[H_1-H_2,E_{\pm \alpha_2}]=\pm  E_{\pm \alpha_2},\nn
&&[H_1+H_2,E_{\pm \alpha_1}]=\pm E_{\pm \alpha_1}~~~,~~~
[H_1+H_2,E_{\pm \alpha_2}]=\mp  E_{\pm \alpha_2},\nn
&&\{ E_{\alpha_1},E_{-\alpha_1}\}=H_2~~~,~~~
\{ E_{\alpha_2},E_{-\alpha_2}\}=H_1~~~,~~~
\{E_{\pm\alpha_1},E_{\pm \alpha_2}\}=E_{\pm(\alpha_1+\alpha_2)}.
\label{eq:cr}\eea
The fermionic nonzero roots $\pm \alpha_1,\pm \alpha_2$ have length 
square zero, and we normalise the bosonic nonzero roots 
$\pm (\alpha_1 +\alpha_2)$ 
to have length square 2. The root diagram can be represented
in a 2-dimensional Minkowski space with the fermionic roots in the
lightlike directions, as in Fig.1.
\vskip .5cm

\setlength{\unitlength}{0.012500in}%
\begingroup\makeatletter\ifx\SetFigFont\undefined
\def\x#1#2#3#4#5#6#7\relax{\def\x{#1#2#3#4#5#6}}%
\expandafter\x\fmtname xxxxxx\relax \def\y{splain}%
\ifx\x\y   
\gdef\SetFigFont#1#2#3{%
  \ifnum #1<17\tiny\else \ifnum #1<20\small\else
  \ifnum #1<24\normalsize\else \ifnum #1<29\large\else
  \ifnum #1<34\Large\else \ifnum #1<41\LARGE\else
     \huge\fi\fi\fi\fi\fi\fi
  \csname #3\endcsname}%
\else
\gdef\SetFigFont#1#2#3{\begingroup
  \count@#1\relax \ifnum 25<\count@\count@25\fi
  \def\x{\endgroup\@setsize\SetFigFont{#2pt}}%
  \expandafter\x
    \csname \romannumeral\the\count@ pt\expandafter\endcsname
    \csname @\romannumeral\the\count@ pt\endcsname
  \csname #3\endcsname}%
\fi
\fi\endgroup
\begin{picture}(151,180)(017,435)
\thicklines
\multiput(180,603)(16.64190,-22.18921){8}{\line( 3,-4){  8.667}}
\put(305,436){\vector( 3,-4){0}}
\put(180,603){\vector(-3, 4){0}}
\multiput(305,603)(-16.64190,-22.18921){8}{\line(-3,-4){  8.667}}
\put(180,436){\vector(-3,-4){0}}
\put(305,603){\vector( 3, 4){0}}
\put(117,520){\vector(-1, 0){  0}}
\put(117,520){\vector( 1, 0){251}}
\put(243,520){\circle*{6}}
\put(315,435){\makebox(0,0)[lb]{\smash{${\bf \alpha_2}$}}}
\put(315,600){\makebox(0,0)[lb]{\smash{${\bf \alpha_1}$}}}
\end{picture}
\vskip 1cm
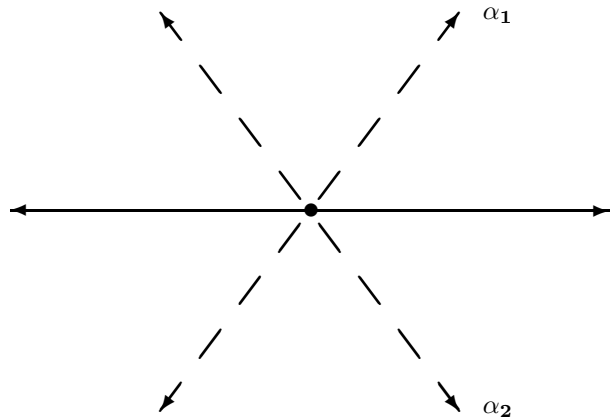
\centerline{{\bf {\rm Fig.1}}:\it ~The root diagram of $A(1,0)$}

\noindent

The Weyl group of $\slr$ is isomorphic to the Weyl group of its even
simple subalgebra $sl(2;{\bf R})$. There is no obvious concept of a
Weyl reflection about the hyperplane orthogonal to a zero square norm
fermionic root.  If one therefore chooses a purely fermionic system of
simple roots $\{\alpha_1,\alpha_2 \}$, there is no element of the Weyl
group which can transform it into the system of simple roots
$\{-\alpha_2,\alpha_1+\alpha_2\}$. Dobrev and Petkova \cite{dp} and 
later, Penkov and Serganova \cite{serga} have actually extended
the definition of the Weyl group to incorporate the transformation
$\alpha_2 \rightarrow -\alpha_2$.  This non uniqueness of
the generalized Dynkin diagram for Lie superalgebras is well
established \cite{kw2}. Let us now determine
the explicit relation between the Wakimoto representations of the
affine version of $\slr$ constructed with both choices of simple
roots.

A standard way to construct a Wakimoto free field representation of
the classical Poisson bracket $\slr$ algebra is to start with a
Wess-Zumino-Witten-Novikov ( WZWN) model based on the noncompact
simple Lie supergroup $\Slr$, introduce a Gauss decomposition for the
supergroup elements, and calculate the currents associated with the
Kac-Moody symmetries of the WZWN action
\cite{ito,K}.  Because
of the non unique choice (up to Weyl
transformations) of the simple roots in $\slr$, any supergroup element
$g$ can be Gauss decomposed in different ways, which lead to different
free field representations.  Our aim is to clarify the relation
between such different representations, both at the classical and at
the quantum level. Let us first summarise the results of \cite{K}.

The WZWN action for the Lie supergroup $\Slr$ in the light cone gauge
is given by
\ben
S(g) = \frac{\kappa}{2} \int dx_+dx_- {\rm Str} (g^{-1}\pa_+g g^{-1} \pa_-g) + 
\kappa \int dx_3 dx_+dx_-               
{\rm Str} (g^{-1}\pa_3g[g^{-1}\pa_+g,g^{-1}\pa_-g]) 
\label{eq:wzw}\een
where the supertrace is the invariant, non-degenerate bilinear form on
$\slr$, and $\kappa$ is a parameter related to the level $k$ of the
Lie superalgebra by the relation $k=-4 \pi \kappa$.  The field
$g(x_+,x_-)$ takes values in the connected real Lie supergroup $\Slr$
and can be parametrised in a neighbourhood of the identity as $g=ABC$
with
\bea
A &=& \exp[ \psi '_1 {\bf e}_{-\alpha_1}+
            \psi '_2 {\bf e}_{-\alpha_2}+
            \gamma {\bf e}_{-(\alpha_1 +\alpha_2)}]\nn
B &=& \exp[\frac{1}{\sqrt{2k}}\fm  {\bf h}_- + \frac{1}{\sqrt{2k}} \fp 
{\bf h}_+]\nn
C &=& \exp[\psi _1 {\bf e}_{\alpha_1}+
           \psi _2 {\bf e}_{\alpha_2}+
            f {\bf e}_{\alpha_1 +\alpha_2}],
\label{eq:gauss1}\eea
or as $g=A'B'C'$ with
\bea
A' &=& \exp[\tilde{\psi} '_1 {\bf e}_{-\alpha_1}+ \tilde{\psi }'_2
{\bf e}_{\alpha_2}+ \tilde{\gamma} {\bf e}_{-(\alpha_1 +\alpha_2)}]\nn
B' &=& \exp[\frac{1}{\sqrt{2k}} \tilde{\phi }_- {\bf h}_-
+\frac{1}{\sqrt{2k}} \tilde{\phi}_+ {\bf h}_+]\nn 
C' &=&
\exp[\tilde{\psi }_1 {\bf e}_{\alpha_1}+ \tilde{\psi }_2 {\bf
e}_{-\alpha_2}+ \tilde{f} {\bf e}_{\alpha_1 +\alpha_2}],
\label{eq:gauss2}\eea
with $\tilde{\psi}'_{1,2}, \tilde{\psi}_{1,2}, \psi'_{1,2}$ and
$\psi_{1,2}$ fermionic parametrisation fields, and $\tilde{\phi}_{1,2},
\tilde{\gamma},\tilde{f}, \phi_{1,2}, \gamma ,f$ 
bosonic parametrisation fields in the variables $x_{\pm}$.  These two
parametrisations, which will be referred to as type I and type II
Gauss decompositions of $g$, single out two inequivalent ways of
choosing the nilpotent (Borel) subalgebras of lowering (factors A and
A') and raising (factors C and C') operators in $\slr$ according to
the choice of simple roots. For instance, for a purely fermionic
choice of simple roots $\{\alpha _1, \alpha_2 \}$, the negative roots are
$\{-\alpha_1,-\alpha_2,-(\alpha_1+\alpha_2)\}$ and the Borel
subalgebra of lowering operators is therefore
$\{E_{-\alpha_1},E_{-\alpha_2},E_{-(\alpha_1+\alpha_2)}\}$. For the
non Weyl-equivalent choice $\{-\alpha _2,
\alpha_1+\alpha_2 \}$, the negative roots are $\{-\alpha_1,\alpha_2,
-(\alpha_1+\alpha_2)\}$ and the corresponding Borel subalgebra is
$\{E_{-\alpha_1},E_{\alpha_2},E_{-(\alpha_1+\alpha_2)}\}$\\

In order to relate the parametrisation fields in the type I and type II
decompositions, it suffices to exploit the equality $ABC=A'B'C'$. 
A tedious but straightforward calculation, using the Hausdorff-Campbell 
formula as well as 
the identity
\ben
\exp[\tpbp {\bf e}_{\alpha_2}] \Bp \exp[\tpsb {\bf e}_{-\alpha_2}]
=
\exp[\alpha {\bf e}_{-\alpha_2}]\Bp \exp[\hf \alpha '\tpsb({\bf h}_-+
{\bf h}_+)]\exp[\alpha '{\bf e}_{\alpha_2}]
\een
with
\bea
\tilde{\phi}\cdot {\bf h}&=& \irt (\tfm {\bf h}_- +\tfp {\bf h}_+)\nn
\alpha = \exp[-\irt (\tfm -\tfp)]\tpsb &,&
\alpha '= \exp[-\irt (\tfm -\tfp)]\tpbp, 
\eea
provides the following classical relations,
\bea
f=\tilde{f}-\hf (\tilde{\psi}_1-\hf \tilde{f}\tilde{\psi}_2)
e^{-\frac{1}{\sqrt{2k}} (\tilde{\phi}_--\tilde{\phi}_+)}
\tilde{\psi}'_2~~~&,&~~~
\gamma = \tilde{\gamma}+\hf (\tilde{\psi}'_1-\hf \tilde{\gamma}
\tilde{\psi}'_2)
e^{-\frac{1}{\sqrt{2k}} (\tilde{\phi}_--\tilde{\phi}_+)}\tilde{\psi}_2\nn
\fm=\tilde{\phi}_-+\sqrt{\frac{k}{2}}
e^{-\frac{1}{\sqrt{2k}} (\tilde{\phi}_--\tilde{\phi}_+)}
\tilde{\psi}'_2\tilde{\psi}_2~~~&,&~~~
\fp= \tilde{\phi}_++\sqrt{\frac{k}{2}}
e^{-\frac{1}{\sqrt{2k}} (\tilde{\phi}_--\tilde{\phi}_+)}
\tilde{\psi}'_2\tilde{\psi}_2\nn
\psi'_1=\tilde{\psi}'_1-\hf\tilde{\gamma}\tilde{\psi}'_2~~~&,&~~~
\psi_1=\tilde{\psi}_1-\hf\tilde{f}\tilde{\psi}_2\nn
\psi'_2= e^{-\frac{1}{\sqrt{2k}} (\tilde{\phi}_--\tilde{\phi}_+)}
\tilde{\psi}_2
{}~~~&,&~~~
\psi_2= e^{-\frac{1}{\sqrt{2k}} (\tilde{\phi}_--\tilde{\phi}_+)}
\tilde{\psi}'_1,
\label{eq:rel}\eea
which are the key to understand the link between the Wakimoto representations
associated to the type I and II decompositions, as we shall now describe.\\

Let us first stress that the relations \eqn{rel} provide canonical 
transformations for the Poisson brackets of the parametrisation fields 
and their conjugate momenta. Indeed, 
when use is made of the Gauss decomposition in the WZWN
action, the Polyakov-Wiegmann identity for $S(ABC)$ \cite{PW} reduces to
\ben
S(ABC)= S(B) + \kappa \int dx_+dx_- 
{\rm Str} (A^{-1}(\pa_- A )B (\pa_+ C)    
   C^{-1}B^{-1}).
\label{eq:polywieg}\een
$S(B)$ is the free field action, which can be calculated in terms of the
parametrisation fields $\fm,\fp $ using \eqn{wzw} and \eqn{gauss1}:
\ben
S(B)= -\frac{1}{8\pi} \int dx_+dx_- 
[\pa_+\fm\pa_-\fm-\pa_+\fp\pa_-\fp],
\een
while the second term in \eqn{polywieg} gives the interaction as 
\bea
\kappa \int dx_+dx_- 
  &\{& e^{\ifrt\fm}[ \pa_-\gamma+ \hf(\pa_-\psi'_2)\psi'_1
-\hf\psi'_2\pa_-\psi'_1]~  
              [\pa_+f-\hf(\pa_+\psi_2)\psi_1+\hf\psi_2\pa_+\psi_1]\nn 
  && +  e^{\frac{1}{\rt}(\fm+\fp)}\pa_-\psi'_1\pa_+\psi_1
       -e^{\frac{1}{\rt}(\fm-\fp)}\pa_-\psi'_2\pa_+\psi_2 \}.
\eea
The classical momenta conjugated to the various parametrisation fields can be
easily derived from the action $S(ABC)$. One has,
\bea
\Pi_{\fm}&=&4\pi \frac{\pa S}{\pa(\pa_-\fm)}=
-\hf \pa _+\fm~~,
{}~~\Pi_{\fp}=4\pi \frac{\pa S}{\pa(\pa_-\fp)}=
\hf \pa _+\fp\nn
\Pi_{\gamma}&=& 4\pi \frac{\pa S}{\pa(\pa_-\gamma)}
=-k e^{\ifrt\fm}(\pa_+f+\hf (\psi_1 \pa_+\psi_2
-(\pa_+\psi_1)\psi_2)\equiv \beta\nn
\Pi_{\psi'_1}&=&4\pi \frac{\pa S}{\pa(\pa_-\psi'_1)}=
\frac{\beta}{2}\psi'_2
               -k e^{\frac{1}{\rt}(\fm+\fp)}\pa_+\psi_1\equiv 
\psi^{'\dagger}_1\nn
\Pi_{\psi'_2}&=&4\pi \frac{\pa S}{\pa(\pa_-\psi'_2)}=
\frac{\beta}{2}\psi'_1
               +k e^{-\frac{1}{\rt}(\fm-\fp)}\pa_+\psi_2\equiv 
\psi^{'\dagger}_2. 
\eea
If one instead starts with $S(A'B'C')$, a very
similar procedure leads to the following conjugate momenta,
\bea
\Pi_{\tilde{\phi}_-}&=&4\pi \frac{\pa S}{\pa(\pa_-\tilde{\phi}_-)}=
-\hf \pa _+\tilde{\phi}_-~~,
{}~~\Pi_{\tilde{\phi}_+}=4\pi \frac{\pa S}{\pa(\pa_-\tilde{\phi}_+)}=
\hf \pa _+\tilde{\phi}_+\nn
\Pi_{\tilde{\gamma}}&=& 4\pi \frac{\pa S}{\pa(\pa_-\tilde{\gamma})}
=-k e^{\ifrt \tilde{\phi}_-}\pa_+\tilde{f}
-\hf \tilde{\psi}^{'\dagger}_1\tilde{\psi}'_2\equiv \tilde{\beta}\nn
\Pi_{\tilde{\psi}'_1}&=& 4\pi \frac{\pa S}{\pa(\pa_-\tilde{\psi}'_1)}=
-k e^{\frac{1}{\rt}(\tilde{\phi}_-+\tilde{\phi}_+)}
(-\hf\tilde{f}\pa_+\tilde{\psi}_2
+\hf \tilde{\psi}_2 \pa_+\tilde{f}+\pa_+\tilde{\psi}_1) 
\equiv \tilde{\psi}^{'\dagger}_1\nn
\Pi_{\tilde{\psi}'_2}&=& 4\pi \frac{\pa S}{\pa(\pa_-\tilde{\psi}'_2)}=
-k e^{-\frac{1}{\rt}(\tilde{\phi}_--\tilde{\phi}_+)}\pa_+\tilde{\psi}_2
-\hf \tilde{\gamma} \tilde{\psi}^{'\dagger}_1 \equiv \tilde{\psi}
^{'\dagger}_2 ,
\eea
and it is easy to obtain the following relations between the type I 
and II conjugate momenta from \eqn{rel},
\bea
\beta &=& \tilde{\beta }-\hf \tilde{\psi}^{'\dagger}_1\tilde{\psi}'_2\nn
\psi^{'\dagger}_1 &=& \tilde{\psi}^{'\dagger}_1-\hf \tilde{\beta}
      e^{-\frac{1}{\sqrt{2k}}(\tilde{\phi}_--\tilde{\phi}_+)}\tilde{\psi}_2
      +\frac{1}{4}e^{-\frac{1}{\sqrt{2k}}(\tilde{\phi}_--\tilde{\phi}_+)}
      \tilde{\psi}^{'\dagger}_1\tilde{\psi}'_2\tilde{\psi}_2\nn
\psi^{'\dagger}_2&=& k\pa_+\tilde{\psi}'_2 -\sqrt{\frac{k}{2}}
    \tilde{\psi}'_2 (\pa_+\tilde{\phi}_--\pa_+\tilde{\phi}_+)
    +\hf \tilde{\beta }\tilde{\psi}'_1- \frac{1}{4}\tilde{\beta}\tilde{\gamma}
    \tilde{\psi}'_2-\frac{1}{4}\tilde{\psi}^{'\dagger}_1\tilde{\psi}'_2
    \tilde{\psi}'_1
\label{eq:rel2}\eea
The fundamental Poisson brackets for the two Gauss decompositions are
taken to be,
\bea
\{ \psi'_i(x),\psi^{'\dagger}_j(y) \}_{P.B.} &=& -\delta_{ij}\delta (x-y)=
\{ \tilde{\psi}'_i(x),\tilde{\psi}^{'\dagger}_j(y) \}_{P.B.}\nn
\{ \gamma(x),\beta (y)\}_{P.B.}&=& \delta(x-y)=
\{ \tilde{\gamma}(x),\tilde{\beta} (y)\}_{P.B.}\nn
\{ \pa_+\phi_a(x),\pa_+\phi_b(y)\}_{P.B.}&=& 
                       \eta _{ab}\delta'(x-y)=
\{ \pa_+\tilde{\phi}_a(x),\pa_+\tilde{\phi}_b(y)\}_{P.B.}\eea
with $\eta _{ab}={\rm diag}(-1,1), a,b=-,+$, 
and they are related by the transformations \eqn{rel} and \eqn{rel2}.\\

The currents associated with the Kac-Moody symmetries of the WZWN action
provide a free field representation of the classical Poisson bracket
algebra $\slr$. Indeed, 
using the Noether method, they can be constructed from the action 
\eqn{wzw} as,
\ben
J(\mbox{\boldmath $\lambda$})=-k{\rm Str}(\mbox{\boldmath $\lambda$}\pa_+g 
g^{-1})
\een
where \mbox{\boldmath $\lambda$} runs over the set of basis matrices (3).
{}For the type I decomposition, the free field representation is given
by ($\pa_+ \equiv \pa $),
\bea
J({\bf h_+})&=&\rt \pa \fp+\papd\pap -\pbpd\pbp\nn
J({\bf h_-})&=&-\rt \pa \fm+\papd\pap +\pbpd\pbp-2\beta \gamma\nn
J({\bf e_{\alpha_1+\alpha_2}})&=& -\rt \gamma \pa \fm +\frt \pap \pbp \pa \fp
  +\frac{k}{2}(\pa \pap \pbp+\pa \pbp \pap)\nn
&&-k\pa \gamma 
  +\gamma (\papd \pap+\pbpd \pbp ) -\beta \gamma ^2\nn
J({\bf e_{-(\alpha_1+\alpha_2)}})&=& \beta\nn
J({\bf e_{\alpha_1}})&=& -\frt \pap(\pa \fm +\pa \fp)-k\pa \pap -\gamma \pbpd 
+\hf \pbpd\pbp\pap -\hf \beta \gamma \pap\nn
J({\bf e_{ -\alpha_1}})&=& \papd+\hf \beta \pbp\nn
J({\bf e_{\alpha_2}})&=& -\frt \pbp(\pa \fm -\pa \fp)-k\pa \pbp -\gamma \papd 
+\hf \papd\pap\pbp -\hf \beta \gamma \pbp\nn
J({\bf e_{-\alpha_2}})&=& -\pbpd-\hf \beta \pap .
\label{eq:typeI}\eea

{}For type II, one gets,
\bea
J({\bf h_+})&=&\rt \pa \tfp+\tpapd\tpap +\tpbpd\tpbp\nn
J({\bf h_-})&=&-\rt \pa \tfm+\tpapd\tpap -\tpbpd\tpbp-2\tilde{\beta} 
\tilde{\gamma}\nn
J({\bf e_{\alpha_1+\alpha_2}})&=& -\rt \tilde{\gamma} \pa \tfm 
  -k\pa \tilde{\gamma} 
  +\hf \tilde{\gamma} (\tpapd \tpap-\tpbpd \tpbp ) 
-\tilde{\beta}\tilde{\gamma} ^2 + \tpap \tpapd-\frac{1}{4}\tilde{\gamma}^2
\tpapd \tpbp \nn
J({\bf e_{-(\alpha_1+\alpha_2)}})&=& \tilde{\beta} -\hf \tpapd \tpbp\nn
J({\bf e_{\alpha_1}})&=& -\frt \tpap(\pa \tfm +\pa \tfp)+\hf \frt \tilde{\gamma}
\tpbp (3 \pa \tfm -\pa \tfp)  -\frac{k}{2}(\tilde{\gamma} \pa \tpbp 
- \pa \tilde{\gamma} \tpbp)\nn
&&-k\pa \tpap+\hf \tilde{\beta} \tilde{\gamma}^2 \tpbp
-\tilde{\beta}\tilde{\gamma} \tpap + \tpbpd \tpap \tpbp\nn
J({\bf e_{-\alpha_1}})&=& \tpapd\nn
J({\bf e_{\alpha_2}})&=& \tpbpd-\hf \tilde{\gamma}\tpapd \nn
J({\bf e_{-\alpha_2}})&=& \frt \tpbp(\pa \tfm -\pa \tfp)-k\pa \tpbp 
-\tilde{\beta} \tpapd 
+\hf \tpapd\tpbp\tpap +\hf \tilde{\beta} \tilde{\gamma} \tpbp.
\label{eq:typeII}\eea
These two free field representations are related through \eqn{rel} and
\eqn{rel2}, as first discussed in \cite{K}.\\

At the quantum level, the free fields become operators whose short distance
behaviour is governed by the following expressions,
\bea
\psi'_i(z)\psi^{'\dagger}_j(w) &\sim& -\frac{\delta_{ij}}{z-w} \sim 
\tilde{\psi}'_i(z)
\tilde{\psi}^{'\dagger}_j (w)\nn 
\gamma (z) \beta (w) &\sim & \frac{1}{z-w} \sim \tilde{\gamma}(z) 
\tilde{\beta}(w)\nn
\pa \phi_a (z) \pa \phi _b (w) & \sim & \frac{\eta_{ab}}{(z-w)^2} \sim 
\pa \tilde{\phi}_a(z) \pa \tilde{\phi}_b (w).
\eea
The "quantum" momenta are very similar to the classical conjugate momenta, 
\bea
\beta &=& -(k+\hf ) e^{-2\alpha_-\fm }(\pa_+f+\hf (\psi_1 \pa_+\psi_2
-(\pa_+\psi_1)\psi_2)\nn
\psi^{'\dagger}_1&=&\frac{\beta}{2}\psi'_2
               -(k +\hf) e^{-\alpha_-(\fm+\fp)}\pa_+\psi_1\nn
\psi^{'\dagger}_2&=& \frac{\beta}{2}\psi'_1
               +(k+\hf) e^{\alpha_-(\fm-\fp)}\pa_+\psi_2\nn 
\tilde{\beta}&=& -(k+\hf ) e^{-2\alpha_- \tilde{\phi}_-}\pa_+\tilde{f}
-\hf \tilde{\psi}^{'\dagger}_1\tilde{\psi}'_2\nn
\tilde{\psi}^{'\dagger}_1&=& -(k+ \hf) e^{-\alpha_-
(\tilde{\phi}_-+\tilde{\phi}_+)}
(-\hf\tilde{f}\pa_+\tilde{\psi}_2
+\hf \tilde{\psi}_2 \pa_+\tilde{f}+\pa_+\tilde{\psi}_1)\nn 
\tilde{\psi}^{'\dagger}_2 &=&
-(k+ \hf ) e^{\alpha_-(\tilde{\phi}_--\tilde{\phi}_+)}\pa_+\tilde{\psi}_2
-\hf \tilde{\gamma} \tilde{\psi}^{'\dagger}_1, 
\eea
where we have defined
\ben
\alpha_-=-i \frac{\sqrt{2k+2}}{2k+1}=-i\frac{\alpha_+}{2k+1},
\een
and where normal ordering of the operators is implicit. We adopt the
"conventional" normal ordering , which implies the following schematic
rule for products of three or more operators,
\bea
O_1O_2O_3&=& {}^{\rm x}_{\rm x}~ O_1~ :~ O_2 O_3 ~:~ {}^{\rm x}_{\rm x}\nn
O_1O_2O_3O_4 &=& \vdots ~O_1 ~{}^{\rm x}_{\rm x} 
{}~O_2 ~:O_3 O_4~:~ {}^{\rm x}_{\rm x}~ \vdots
\label{eq:nop}\eea 
The affine $ \hslr$ Lie
superalgebra is defined through the following operator product expansions
between currents,
\bea
J({\bf e}_{\alpha_1+\alpha_2})(z)~J({\bf e}_{-(\alpha_1+\alpha_2)})(w)
& \sim & \frac{J({\bf h}_-)(w)}{z-w} +\frac{k}{(z-w)^2}\nn
J({\bf h}_-)(z)~J({\bf e}_{\pm(\alpha_1+\alpha_2)})(w) &\sim &
\pm \frac{2J({\bf e}_{\pm(\alpha_1+\alpha_2)})(w)}{z-w}\nn
J({\bf e}_{\pm(\alpha_1+\alpha_2)})(z)~J({\bf e_{\mp \alpha_1}})(w) & \sim &
\pm \frac{J({\bf e_{\pm \alpha_2}})(w)}{z-w}\nn
J({\bf e}_{\pm(\alpha_1+\alpha_2)})(z)~J({\bf e_{\mp \alpha_2}})(w) & \sim &
\mp \frac{J({\bf e_{\pm \alpha_1}})(w)}{z-w}\nn
J({\bf h}_-)(z)~J({\bf e_{\pm \alpha_1}})(w) \sim 
\pm \frac{J({\bf e_{\pm \alpha_1}})(w)}{z-w}~~~&,&~~~
J({\bf h}_-)(z)~J({\bf e_{\pm \alpha_2}})(w)\sim
\pm \frac{J({\bf e_{\pm \alpha_2}})(w)}{z-w}\nn
J({\bf h}_+)(z)~J({\bf e_{\pm \alpha_1}})(w) \sim  
\pm \frac{J({\bf e_{\pm \alpha_1}})(w)}{z-w}~~~&,&~~~
J({\bf h}_+)(z)~J({\bf e_{\pm \alpha_2}})(w) \sim 
\mp \frac{J({\bf e_{\pm \alpha_2}})(w)}{z-w}\nn
J({\bf h}_{\pm})(z)~J({\bf h}_{\pm})(w)& \sim &
\mp \frac{k}{(z-w)^2}\nn
J({\bf e_{\alpha_1}})(z)~J({\bf e_{-\alpha_1}})(w)& \sim &
\hf \frac{(-J({\bf h}_-)+J({\bf h}_+)(w))}{z-w}-\frac{k}{(z-w)^2}\nn
J({\bf e_{\alpha_2}})(z)~J({\bf e_{-\alpha_2}})(w)& \sim &
\hf \frac{(J({\bf h}_-)+J({\bf h}_+))(w)}{z-w}+\frac{k}{(z-w)^2}\nn
J({\bf e_{\pm \alpha_1}})(z)~J({\bf e_{\pm \alpha_2}})(w)& \sim &
\frac{J({\bf e}_{\pm(\alpha_1+\alpha_2)})(w)}{z-w}
\eea
As in the classical case, there exist two Wakimoto representations of this
affine Lie superalgebra, which we will denote type I and type II since their
classical limit (obtained when taking the level $k$ to the $\infty $ limit)
coincides with the type I and II representations \eqn{typeI},\eqn{typeII}.
The type I representation has been given by \cite{bershad} in their discussion
of Hamiltonian reduction of the affine version of $Osp(2/2)$, and it appears
as well in \cite{ito} and \cite{kimura}. We shall include it here for 
completeness, and to allow direct comparison with the type II representation
which we have not found in the literature. The type I free field representation
is, with implicit conventional normal ordering,
\bea
J({\bf h_+})&=&\iap \pa \fp+\papd\pap -\pbpd\pbp\nn
J({\bf h_-})&=&-\iap \pa \fm+\papd\pap +\pbpd\pbp-2\beta \gamma\nn
J({\bf e_{\alpha_1+\alpha_2}})&=& -\iap \gamma \pa \fm +\hf \iap \pap \pbp 
\pa \fp
  +\hf (k+1)(\pa \pap \pbp+\pa \pbp \pap)\nn
&&-k\pa \gamma 
  +\gamma (\papd \pap+\pbpd \pbp ) -\beta \gamma ^2\nn
J({\bf e_{-(\alpha_1+\alpha_2)}})&=& \beta\nn
J({\bf e_{\alpha_1}})&=& -\hf \iap \pap(\pa \fm +\pa \fp)-\hf (2k+1)\pa 
\pap -\gamma \pbpd 
+\hf \pbpd\pbp\pap -\hf \beta \gamma \pap\nn
J({\bf e_{ -\alpha_1}})&=& \papd+\hf \beta \pbp\nn
J({\bf e_{\alpha_2}})&=& -\hf \iap \pbp(\pa \fm -\pa \fp)-\hf (2k+1)\pa 
\pbp -\gamma \papd 
+\hf \papd\pap\pbp -\hf \beta \gamma \pbp\nn
J({\bf e_{-\alpha_2}})&=& -\pbpd-\hf \beta \pap .
\label{eq:qtypeI}\eea
{}For type II, one gets,
\bea
J({\bf h_+})&=&\iap \pa \tfp+\tpapd\tpap +\tpbpd\tpbp\nn
J({\bf h_-})&=&-\iap \pa \tfm+\tpapd\tpap -\tpbpd\tpbp-2\tilde{\beta} 
\tilde{\gamma}\nn
J({\bf e_{\alpha_1+\alpha_2}})&=& -\iap \tilde{\gamma} \pa \tfm 
  -(k+\hf)\pa \tilde{\gamma} 
  +\hf \tilde{\gamma} (\tpapd \tpap-\tpbpd \tpbp ) 
-\tilde{\beta}\tilde{\gamma} ^2 + \tpap \tpapd-\frac{1}{4}\tilde{\gamma}^2
\tpapd \tpbp \nn
J({\bf e_{-(\alpha_1+\alpha_2)}})&=& \tilde{\beta} -\hf \tpapd \tpbp\nn
J({\bf e_{\alpha_1}})&=& -\hf \iap \tpap(\pa \tfm +\pa \tfp)+\frac{1}{4} 
\iap \tilde{\gamma}
\tpbp (3 \pa \tfm -\pa \tfp)  -\hf (k+1)\tilde{\gamma} \pa \tpbp \nn
&&+\hf (k-1) \pa \tilde{\gamma} \tpbp)
-k\pa \tpap+\hf \tilde{\beta} \tilde{\gamma}^2 \tpbp
-\tilde{\beta}\tilde{\gamma} \tpap + \tpbpd \tpap \tpbp\nn
J({\bf e_{-\alpha_1}})&=& \tpapd\nn
J({\bf e_{\alpha_2}})&=& \tpbpd-\hf \tilde{\gamma}\tpapd \nn
J({\bf e_{-\alpha_2}})&=& \hf \iap  \tpbp(\pa \tfm -\pa \tfp)
-\hf (2k+1)\pa \tpbp -\tilde{\beta} \tpapd 
+\hf \tpapd\tpbp\tpap +\hf \tilde{\beta} \tilde{\gamma} \tpbp,\nn
\label{eq:qtypeII}\eea
where
\ben
\alpha_+ = \sqrt{2k+2}.
\een
Having explicitly provided two genuinely different Wakimoto free field 
representations of $\hslr$, whose existence is intimately rooted in the
non-uniqueness of the Dynkin diagram of $\slr$, it is useful to
identify the relations between the free fields of these two representations.
They read,  
\bea
\gamma &=& \tilde{\gamma}+\hf (\tilde{\psi}'_1-\hf \tilde{\gamma}
\tilde{\psi}'_2)
e^{\alpha_- (\tilde{\phi}_--\tilde{\phi}_+)}\tilde{\psi}_2\nn
\fm &=& \tilde{\phi}_--\frac{1}{2\alpha_-}
e^{\alpha_- (\tilde{\phi}_--\tilde{\phi}_+)}
\tilde{\psi}'_2\tilde{\psi}_2~~~,~~~
\fp = \tilde{\phi}_+-\frac{1}{2\alpha_-}
e^{\alpha_- (\tilde{\phi}_--\tilde{\phi}_+)}
\tilde{\psi}'_2\tilde{\psi}_2\nn
\psi'_1&=&\tilde{\psi}'_1-\hf \tilde{\gamma}\tilde{\psi}'_2~~~ ,~~~
\psi'_2= e^{\alpha_-(\tilde{\phi}_--\tilde{\phi}_+)}
\tilde{\psi}_2\nn
\beta &=& \tilde{\beta }-\hf \tilde{\psi}^{'\dagger}_1\tilde{\psi}'_2\nn
\psi^{'\dagger}_1 &=& \tilde{\psi}^{'\dagger}_1-\hf \tilde{\beta}
      e^{\alpha_-(\tilde{\phi}_--\tilde{\phi}_+)}\tilde{\psi}_2
      +\frac{1}{4}e^{\alpha_-(\tilde{\phi}_--\tilde{\phi}_+)}
      \tilde{\psi}^{'\dagger}_1\tilde{\psi}'_2\tilde{\psi}_2\nn
\psi^{'\dagger}_2&=& (k+\hf)\pa_+\tilde{\psi}'_2 -\hf \iap
    \tilde{\psi}'_2 (\pa_+\tilde{\phi}_--\pa_+\tilde{\phi}_+)
    +\hf \tilde{\beta }\tilde{\psi}'_1- \frac{1}{4}\tilde{\beta}\tilde{\gamma}
    \tilde{\psi}'_2-\frac{1}{4}\tilde{\psi}^{'\dagger}_1\tilde{\psi}'_2
    \tilde{\psi}'_1\nn\label{eq:qrel}\eea

It can be checked that substituting these relations into the type I
currents \eqn{qtypeI} yields the type II expressions \eqn{qtypeII},
provided the normal ordering defined in \eqn{nop} is carefully implemented.
In order to calculate the BRST cohomology on  the whole Fock space of 
the \break
$\Slr/\Slr$ topological model, one needs the Wakimoto representation 
\eqn{qtypeI} and its dual. The two are related by the following automorphism
of order 4 of the $\hslr$ algebra,

\bea
J^{\pm} \rightarrow -J^{\mp},&&~~~J^3 \rightarrow -J^3,~~~U \rightarrow -U\nn
j^{\pm} \rightarrow \pm j^{\mp},
&&~~~j'^{\pm} \rightarrow \mp j'^{\mp},~~~k \rightarrow -(k+2).
\eea

To conclude, we stress  that in order to study the space
of physical states of the noncritical $N=2$ theory, using the tool provided by 
topological
$G/G$ WZNW models, a detailed analysis of various modules over $A(1,0)^{(1)}$
is needed.
Many Lie superalgebras share with $A(1,0)$ the property that two sets of 
simple roots may not be equivalent up to Weyl tranformations, which are 
generated by reflections with respect to bosonic simple roots. An added 
technical complication in $A(1,0)$ is the fact that the fermionic roots are 
lightlike,
which prevents one from defining coroots and fundamental weights in a
straightforward way. We have given
the classical and quantum free field Wakimoto representations of $\hslr$ 
and shown that two Wakimoto representations 
built with two inequivalent
sets of simple roots  are different. Classically, the relation between the two
can be derived from first principles. Remarkably, there also exists
a set of field transformations which relate the two Wakimoto representations
in the quantum case.\\

{\bf Acknowledgements}

Anne Taormina acknowledges the U.K. Engineering and Physical Sciences
Research Council for the award of an Advanced Fellowship.


\begin{thebibliography}{99}
\bibitem{goulian} Goulian, M. and Li, M.,
                  Phys. Rev. Lett. {\bf 66} (1991) 2051.
\bibitem{distler} Distler, J., Hlousek, Z. and  Kawai, H.,
              Int. J. Mod. Phys. {\bf A5} (1990) 391.
\bibitem{fy2} Fan, J-B. and Yu, M.,  
              Preprint AS-ITP-93-22 (1993), hep-th 9304122. 
\bibitem{abdalla1} Abdalla, E. and Zadra, A.,                    
                   Nucl. Phys. B{\bf 432} (1994) 163. 
\bibitem{anton} Antoniadis, I., Bachas C. and Kounnas, C.,                
                 Phys. Lett. {\bf 242,2} (1990) 185.
\bibitem{abdalla2} Abdalla, E., Abdalla, M.C.B., and Dalmazi, D.,
                   Phys. Lett. {\bf 291} (1992) 32.
\bibitem{yankiel} Aharony, G., Ganor, O., Sonnenschein, J., Yankielowicz, S.
                  and Sochen, N.,                 
                  Nucl. Phys. B{\bf 399} (1993) 527.
\bibitem{huyu} Hu, H.L. and Yu, M., 
                Phys. Lett. B{\bf 289} (1992) 302; 
                Nucl. Phys. B{\bf 391} (1993) 389.
\bibitem{fy1} Fan, J-B. and Yu, M.,
              Preprint AS-ITP-93-14 (1993), hep-th 9304122.
\bibitem{us} Bowcock, P. and Taormina, A., Noncritical $N=2$ strings,
             to appear.
\bibitem{yankiel2} Aharony, G., Sonnenschein, and J., Yankielowicz, S.,                             
                   Phys. Lett. B{\bf 289} (1992) 309.
\bibitem{sadov} Sadov, V.,
                Int. J. Mod. Phys. {\bf A8} (1993) 5115.
\bibitem{jens} Petersen, J.L., Rasmussen, J. and Yu, M.,               
               Nucl. Phys. B{\bf 457} (1995) 309.
\bibitem{bouw} Bouwknegt, P., Mc Carthy, J. and Pilch, K.,                
               Comm. math. Phys. {\bf 145} (1992) 541.
\bibitem{PW} Polyakov, A.M. and Wiegmann, P.B.,              
             Phys. Lett. B{\bf 131} (1983) 121;
             Phys. Lett. B{\bf 141} (1984) 223.
\bibitem{ks} Karabali, D. and Schnitzer, H.J.,              
             Nucl. Phys. B{\bf 329} (1990) 625.
\bibitem{gk} Gawedski, K. and Kupiainen A.,              
             Nucl. Phys. B{\bf 320} (1989) 649.
\bibitem{kac77} Kac, V.G. ,                
                Adv. Math. {\bf 26} (1977) 8.
\bibitem{kw} Kac, V.G.and Wakimoto M.,              
             Proc. Nat. Acad. Sci. {\bf 85} (1988) 4956.
\bibitem{us2} Bowcock P. and Taormina A., Representation theory of the affine
              Lie superalgebra $\hslc$ at fractional level, hepth 9605220.
\bibitem{dp}  Dobrev, V.K. and Petkova, V.B., 
              Fortschr. Phys. {\bf 35, 7} (1987) 537.
\bibitem{serga} Penkov, I. and Serganova, V.,
                Indag. Math. {\bf  N.S.3(4)} (1992) 419.
\bibitem{kw2} Kac, V.G. and Wakimoto, M.,  
              hep-th 9407057 (1994).
\bibitem{ito} Ito, K.,              
              Phys. Lett. {\bf B259} (1991) 73. 
\bibitem{K} Koktava, R-L. K.,                               
            Phys. Lett. B{\bf351} (1995) 476.
\bibitem{bershad} Bershadsky,M.and Ooguri, H.,
                  Phys. Lett. {\bf B229} (1989) 374.
\bibitem {kimura} Kimura, K., 
                  Int. Journ. Mod. Phys. {\bf A7}, suppl.1B (1992) 533.
\end{thebibliography}
\end{document}